# Inferential statistics, power estimates, and study design formalities continue to suppress biomedical innovation

**Scott E. Kern**
The Sidney Kimmel Comprehensive Cancer Center at Johns Hopkins, Dept. of Oncology, 1650 Orleans St, Baltimore, MD 21287, 410-614-3314, sk@jhmi.edu.

Supported by NIH grants CA62924 and 134292 and by the Everett and Marjorie Kovler Professorship in Pancreas Cancer Research.

*Abstract*

Innovation is the direct intended product of certain styles in research, but not of others. Fundamental conflicts between descriptive vs inferential statistics, deductive vs inductive hypothesis testing, and exploratory vs pre-planned confirmatory research designs have been played out over decades, with winners and losers and consequences. Longstanding warnings from both academics and research-funding interests have failed to influence effectively the course of these battles. The NIH publicly studied and diagnosed important aspects of the problem a decade ago, resulting in outward changes in the grant review process but not a definitive correction. Specific reforms could deliberately abate the damage produced by the current overemphasis on inferential statistics, power estimates, and prescriptive study design. Such reform would permit a reallocation of resources to historically productive rapid exploratory efforts and considerably increase the chances for higher-impact research discoveries. We can profit from the history and foundation of these conflicts to make specific recommendations for administrative objectives and the process of peer review in decisions regarding research funding. © 2013 S. Kern

"There is nothing more necessary to the man of science than its history, and the logic of discovery…: the way error is detected, the use of hypothesis, of imagination, the mode of testing." – Lord Acton, quoted by Karl Popper (2)

"The most striking feature of the normal research problems we have just encountered is how little they aim to produce major novelties, conceptual or phenomenal…everything but the most esoteric detail is known in advance, and the typical latitude of expectation is only somewhat wider…Normal science does not aim at novelties of fact or theory and, when successful, finds none." – Thomas Kuhn (4)

A pessimist, an optimist, an inferential statistician, and a descriptive statistician go into a bar. They order beers for everyone. When their own drinks arrive, the pessimist complains that his glass came half empty. The optimist expresses begrudging satisfaction that his is at least half full. The inferential statistician explains that one cannot exclude the null hypothesis, which holds that the half-full and half-empty glasses have been shorted by the same amount of beer. The descriptive statistician shakes his head, explaining that he saw the bartender switch to larger glasses after he had used all of the others.

Academic research productivity is a subject of active research and discussion. Among the major determinants of research productivity is the research mix: the proportions of research devoted to novelty, incremental knowledge, or confirmatory research (5). It thus becomes critical to examine whether the objectives of biomedical research innovation are optimally served by current practices. This line of analysis leads through firmly established historical battlegrounds of publication and funding that remain crucial today. We must examine the key schisms in scientific and statistical philosophy, revisiting the fundamental questions of interest to Kuhn and Popper, Pearson and Tukey. We can then examine the administrative principles governing research policy decisions and consider specific recommendations to rebalance the research mix towards a specific goal of driving biomedical innovation.

## The two branches of statistics

### What are statistics?

Statistics are numerical or graphical summaries of a sample, or group of subjects. A similar summary, characterizing an entire population, is termed a parameter, but parameters are seldom measured in biomedical research. Statistical analysis uses summary numbers by organizing them, observing them, and/or inferring their relationships to each other. Almost any summary statement concerning the information in a database is likely to contain a statistic. A database of primary data, however, is not a statistic. "Statistics" also refers to a group of methods providing the analysis.

### Descriptive and inferential statistics differ

Descriptive and inferential statistics are the two major phyla of the statistical kingdom of mathematics. It is essential to explore the difference in some detail. The difference serves as a foundation for analyzing problems in the research enterprise and for recommending changes in research policies.

Overlap exists, and neither discipline is ignorant of the other. For example, an intuitive understanding of an association can be described using only the observed values (e.g., imagine that all 300 patients with the acquired immunodeficiency syndrome were found to have the AIDS virus, and a corroborative study in a sample of 500 additional similar patients found no exceptions to this pattern of association). Or, one could statistically infer a particular likelihood after observing an initial sample (e.g., after the first study, one could calculate the high numerical likelihood that more than 450 patients in the second sample would be found to have the AIDS virus. This prediction would be corroborated by the second study). These essential similarities are not the subject of the discussion below. Here, I will focus on, and thus exaggerate somewhat for illustration, their distinct tendencies by which they pull research in different directions. (*Please see the boxed text for a tutorial on the differences between Descriptive Statistics (DS) and Inferential Statistics (IS).*)

Considerable differences between the phyla exist and have been at center of a broad philosophical war since at least the 1930s (discussed below). The current predominance of inferential statistics, and specifically the methods intended to test a particular hypothesis, owes its success to an unfortunately confluence of features, practices, and history (6). They include confusion, mis-teachings, fears of sanctions from editors and others, and a capturing of the treasury – referring to the NIH and the strings of funding for biomedical science.

## The historical emergence of inferential statistics

### In the beginning

The medical literature arose as case reports and as larger studies reported using descriptive statistics. Profound qualitative findings were made. Once these associations were noted, few numbers or graphs were needed to teach the resulting rules of clinical practice. Examples follow.

> It was noted that an absence of brain activity associated uniformly with lack of recovery and eventual whole-body death. A still-pumping heart associated with both of two contrasting conditions: a functioning brain and a nonfunctional one. End-of-life clinical decisions are made upon such simple associations. A legal declaration of death can be based on an assessment of brain activity, even when the heart is beating.

> The symptoms of acute pulmonary and cardiovascular collapse were associated with thrombus within the pulmonary artery and with deep venous thrombosis. Based on these associations, the principle emerged of anticoagulant treatment given presumptively, prior to establishing a final diagnosis.

> Metastatic colorectal carcinoma was associated with polyps (i.e., adenomas) having invasion through the muscularis mucosae, but not associated with adenomas lacking invasion. Some patients can accurately be informed, "We are sure that we got it in time", based on such simple associations.

A detailed example is also instructive. When testing the toxicity of imatinib in cell culture, effectiveness at a low concentration (1/10 to 1/20 of the typical toxic concentration) was associated with leukemic cells having the BCR-ABL translocation, but not with cells lacking the translocation (7): This was a descriptive association. When treating patients using imatinib, success was associated with the neoplasms having the BCR-ABL translocation or a KIT gene mutation, and not with other conditions: This was a descriptive association. In recent-diagnosis BCR-ABL-positive leukemia, imatinib therapy was generally preferred over bone marrow transplantation, and this preference could have been in theory proved (i.e., statistically supported) by a randomized trial comparing the two therapies: This would have been an inferential association. The testing of this hypothesis has not yet been done, because bone marrow transplantation itself induces mortality in nearly 5%, while imatinib has not been associated with early therapy-induced death: This discrepancy, and the ethical and common-sense barriers to such a clinical trial, was observed from descriptive statistics. Inference testing may someday establish the best situation for use of marrow transplant in this setting, but the descriptive statistics will set the acceptable boundaries for such a study.

**Starting in about the 1920s**

The basic textbooks in most medical subjects are filled predominantly with time-tested facts obtained by descriptive statistics. Even in psychology, Piaget and Pavlov did not rely on inferential statistics for their major discoveries, and Skinner often criticized their use as well (6). Starting in about the 1920s, the subtle methods of psychology, where effect sizes were often small, welcomed new hypersensitive methods to generate discoveries of causal relationships. The practice spread to the rest of biomedical science. Especially when choosing among alternative therapies, it is often considered valuable to know whether one therapy might be slightly better than another. Thus, in clinical therapeutic trials, inferential statistics are generally used when possible.

**Eventual domination by statistical hypothesis inference testing**

Most biomedical papers cite a p value or depend upon studies having used it, and therefore they use inferential statistics. Many also present counts and averages, and they therefore use descriptive statistics as well. Yet, the former is the zeitgeist of our times. The new norm is an expectation that all biomedical science will be planned, funded, performed, and reported using inferential statistics. Even when a study of simple causal relationships is intended to be exploratory and descriptive, the effort can unfortunately be coerced into the mold of inferential process.

Experts have documented this pattern of domination. Gigerenzer wrote that textbooks and curricula "almost never teach the statistical toolbox, which contains tools such as descriptive statistics, Tukey's exploratory methods, Bayesian statistics, Neyman-Pearson decision theory and Wald's sequential analysis" (6). Campbell, in his in outgoing remarks as editor of leading psychology journal (8), lamented, "One of the most frustrating aspects of the journal business is the null hypothesis. It just will not go away. Books have been written [but] it is almost impossible to drag authors away from their p values...It is not uncommon for over half the space in a results section to be composed of parentheses inside of which are test statistics, degrees of freedom, and p values…Investigators must learn to argue for the significance of their results without reference to inferential statistics."

---

*Descriptive vs inferential statistics: A tutorial*

**Definitions**

Descriptive statistics (DS) organizes and summarizes the observations made. It satisfies the broad curiosity driving an ongoing study.

Inferential statistics (IS) attempts to create conclusions that reach beyond the data observed. It satisfies specific questions raised prior to the study.

**Their goals differ**

DS has a low reliance on starting premises and permits a quick survey to identify high-magnitude patterns. DS can observe new areas of interest as well as pre-existing ones. In a subject area, DS is initially used for exploration; in later stages, it serves to corroborate, or sometimes can disprove by a single counterexample. Thus, intuitive predictions are broadly enabled by DS. It detects qualitative or quantitative differences when they form obviously distinct patterns.

In DS, the conclusions are equivalent to the findings. Conclusions are observed. No interpretive errors are possible. In DS, any "significance" is recognized from familiar and intuitive rules of logic. DS has low capability to detect differences of low magnitude; proposals to use DS indicate minimal desire to uncover subtle findings.

DS examines even small and unanticipated new categories; categories are not typically subject to much manipulation. To encounter biases is expected in early explorations; indeed, their recognition may represent a goal of the research.

DS describes as many diverse characteristics as possible. Individuals can be grouped to create a logical organization of data, but DS also displays individuals freely, in addition to any categories that are used. Thus, DS also detects individuals that are exceptions to the larger patterns.

IS is largely performed by statistical hypothesis inference-testing (terminology suggested by Cohen)(1), a large component of which in turn is null-hypothesis testing (NHT). Bayesian statistics is a separate subset of IS. IS infers the likely patterns governing the data and answers particular numerical questions established before the study began.

(Continued)

**How it happened**

Karl Pearson in 1914 published a table of calculated values of probability ("P") for various random samplings from a population (9). Soon afterwards, in the 1920s, Ronald Fisher introduced a method by which a P value could be used to decide to reject a null hypothesis (10). Even as late as 1955 (6), Fisher's writing about the method still envisioned a null hypothesis lacking a specified statistical alternative hypothesis and omitting the concept of effect sizes.

Jerzy Neyman and Egon Pearson (Karl's son) in 1933 introduced the use of two hypotheses, with a decision criterion to choose among them. They also criticized Fisher's method, in part because of his omission of alternative hypotheses. The hybrid of Neyman's and Pearson's ideas with Fisher's created the modern mode of null-hypothesis testing, although many have written that the two theories are logically incompatible (6, 11). Goodman wryly noted that by trying to supplant Fisher's P value, Neyman and Pearson unintentionally immortalized it (11).

These decision-based methods are distinct from certain other respected theories of statistical analysis, such as descriptive statistics, Tukey's exploratory data analysis, and Bayesian statistics. As Tukey stated, "If one technique of data analysis were to be exalted above all others for its ability to be revealing to the mind in connection with each of many different models, there is little doubt which one would be chosen. The simple graph has brought more information to the data analyst's mind than any other device. It specializes in providing indications of unexpected phenomena" (12). And in Bayesian statistics, the probability of a hypothesis becomes altered by the results of an experiment or observation; a decision is not inherent to the process. Indeed, Goodman noted that when clinical trials are re-analyzed by Bayesian methods, the initially observed differences can often be observed to be untrue (11).

Mis-teaching helped spread the new inferential methods. According to Gigerenzer (6), the most widely read textbook in the subject in the 1940s and 1950s was Guilford's Fundamental Statistics in Psychology and Education. It contained false statements, such as "If the results comes out one way, the hypothesis is probably

---

There is a high reliance on the starting premises of the study. IS is used as an intentionally slow roadblock in research so that a subtle difference can be appropriately vetted before being adopted as true. Typically, IS aims to disprove a "null" hypothesis ("$H_o$") established from earlier descriptive studies or theoretical prediction, termed NHT. Less frequently, IS aims to decide between hypotheses competing on an even basis. IS is intended to predict an outcome or estimate a frequency within quantified limits of accuracy. It is usually used to test quantitative differences.

In IS, the conclusions are extracted from the findings using the given premises and inferences. Interpretive errors are possible. Conclusions are decisions. Low-magnitude differences can earn attention when deemed "statistically significant". "Significance" in IS arises from a numerical result to which is applied a decision rule. Familiar rules of logic may or may not be applicable. IS offers the power to detect differences of low magnitude.

Pre-defined categories are the substrate for IS, categories that have a sufficient size. Ideally, the categories are defined prior to study design and data collection. The category assignments are intended to be free of bias owing to prior characterization (i.e., familiarity with the categories) and study design (such as randomization of treatment assignments to remove biases in a clinical therapeutic trial).

IS detects characteristics that differ between groups, or that fail to differ adequately. Individuals that differ can be detected, but this is usually not the goal. IS will group individuals specifically in order to gain adequate statistical power for a comparison.

**Their fields of use characteristically differ**

Descriptive statistics dominates naturally in the fields of biochemistry, anatomy and developmental biology, molecular and genetic pathology, and when reporting events such as accident rates and injuries. DS is a first choice for the qualitative interpretation of model systems, such as studies of large deletion mutations in proteins, and transgenic gene-knockout cells and animals. DS is highly efficient when developing new technical methods.

Inferential statistics dominates naturally when comparing therapies, comparing groups for their predicted risks, and in behavioral, environmental, and genetic epidemiology, including population studies and pedigree analysis. IS is ideal for quantitative interpretation of model systems in which subtle changes are sought – in any field of study. IS is a valid choice for comparing competing technical methods once developed, if it is desired to characterize minor differences.

**The perspective from cognitive evolution**

Descriptive statistics uses categorical data closely suited for use in the minute-by-minute activities of the human brain. Using DS in a study is analogous to the highly patterned decision-making behaviors employed while an animal body is in motion. Confusion and ambiguity are unwelcome.

Inferential statistics uses subtle numerical distinctions tied to abstract thought. IS is characteristically performed by the human brain when in quiet undistracted contemplation. Confusion and ambiguity can be reasoned away or compartmentalized.

(Continued)

correct, if it comes out another way, the hypothesis is probably wrong." And according to Bakan (13), Fisher in 1947 falsely stated that his principles were "common to all experimentation".

At some point, it became common to refer to a "statistical association" solely according to its definition in inferential statistics. It also became common to refer to "predictive statistics" as synonymous with inferential statistics. Both, however, are utilities also provided by descriptive statistics.

<u>Confusion</u> played its part. Gigerenzer suggested that the typical pattern by which hypotheses are tested, irrespective of the many statistical alternatives available, is a ritual that requires confusion for its propagation (6).

Haller and Krauss posed questions to 113 subjects comprising statistics teachers (including professors, lecturer, teaching assistants) and non-statistics teachers of psychology, and psychology students. The questionnaire contained six false statements about what could be concluded from a p value. None of the students noticed that all statements were wrong. They had apparently learned well from their teachers, because 80% of the statistics teachers and 90% of the other teachers answered that at least one of the statements was true (6). Similar results were also obtained in a separate studies by Oakes (6) and by Rosenthal and Gaito (13).

<u>Fear of sanctions</u> enforced the emerging domination. Gigerenzer told the story of an author of a noted statistical textbook, whose textbook initially informed readers of alternative methods of statistical analysis, but reverted in later editions to a single-recipe approach of hypothesis-testing by p values. When answering Gigerenzer's question of why, the author pointed to "three culprits: his fellow researchers, the university administration, and his publisher." The author himself was "a Bayesian", and yet had deleted the chapter on Bayesian statistics. Gigerenzer summarized, "He had sacrificed his intellectual integrity for success" (6).

Gigerenzer also cited Geoffrey Loftus, editor of Memory and Cognition. Upon taking the editorial position, he encouraged the use of descriptive statistics and did not demand null-hypothesis testing or decisions on hypotheses. Under Loftus, alas, only 6% of articles presented descriptive information without any null-hypothesis testing. Yet, the proportion of articles exclusively relying on the testing of the null hypothesis decreased from 53% to 32%, and it rose again to 55% when he was replaced by another editor who emphasized

---

**Study planning differs**

In <u>descriptive statistics</u>, planning can often be provisional and intuitive. DS is suitable for an experienced investigator to carry out on-the-fly. A null hypothesis is not required, although imprecise hypothetical ideas may properly guide the studies. Research planning revolves around the pre-analytic steps: understanding of the question and the likely dominant variable(s), sample availability and its annotated information, and a toolbox of valid assays. Whether a study should be worthwhile revolves on simple concepts of potential impact, novelty of the subject, feasibility, momentum, and investigator experience.

In <u>inferential statistics</u>, study planning is not always intuitive and is seldom brief. The necessity for a pre-study plan precludes performance on-the-fly. A definitive null hypothesis is routine. Alternative hypotheses may be inferred if not explicitly stated. The research plan includes lists of pre-analytic and post-analytic techniques including the precise premises, the study design, dominant variable(s), and power estimates. It can be difficult to judge at the time of planning whether an experienced investigator has correctly designed a study so as to survive post-publication criticism, due to the many modes of failure potentially threatening the study.

**Methods of displaying results differ**

The strictly numerical summary statistics of <u>descriptive statistics</u> can include: counts of subjects, events, and characteristics; the mean; measures of spread (standard deviation, quartiles, and confidence intervals relating the observed variation and mean) and skew; sensitivity and specificity of associations, odds ratio, simple linear regression, hierarchical clustering, principal components analysis, and tables and arrays of statistics.

The strictly numerical summary statistics of <u>inferential statistics</u> can include: estimated differences in the means or variance of compared groups; confidence intervals used to predict future data; p values either uncorrected or corrected for multiple comparisons; estimated effect sizes, hazard ratios; correlation; statistical comparison tests based on lifetable analysis; multivariate risk analysis; and tables and arrays of such statistics.

The graphical displays of <u>descriptive statistics</u> focus on the empirical distribution and can include the: scatter/dot/stem-leaf plot; box and whisker plot; histogram with error bars; Venn diagram for simple, multivariate, or multidimensional data; time-based plot of observed and complete survival data; ROC (receiver operating characteristic) curve; multidimensional maps; and highly clever visual displays. (Some descriptive organizations of data, such as a scatter plot, actually represent pure data. Their primary purpose is to allow the eye to see the patterns.)

The graphical displays of <u>inferential statistics</u> use idealized depictions derived from the real or postulated data, including the: power estimate curves for planning future studies; Kaplan-Meier estimator (a plotted survival curve) for observed populations having incomplete or censored followup data, or plot of predicted survival curves; and chart annotation to denote which of the displayed differences are statistically significant.

(Continued)

the usual inferential statistical tests (6).

Neurosis. Gigerenzer argued that hypothesis testing was a form of personal subconscious conflict resolution, analogous to a Freudian repetitive behavior, making it resistant to logical arguments (6).

Control of the money. The NIH is the major centralized funder for publicly supported biomedical research in the United States. The entanglement of inferential statistics with the policy-deciding administrative centers of the NIH is discussed below.

**Clarifying caveats**

The proper use of inferential statistics is not in question, whether in the writings of experts or here. Inferential statistics are necessary, for example, to conduct multivariate analysis of co-existing risk factors in epidemiology, to compare treatments in settings where subtle improvements are valued, and to aid hypothesis-testing study designs by using power estimates, which judge the number of subjects or assess the quality of pedigrees required to generate usable conclusions. The inferential power of Bayesian statistics is often appropriate when a hypothesis is being examined, although it is not the dominant strain of inferential statistics published today.

The use of falsification as a deductive tool is also not in question. Both descriptive and inferential statistics are capable of providing scientific falsification of hypothetical alternatives.

The problem under discussion is not that inferential statistics and hypothesis testing are employed, nor that they are employed too often. The problem is that their use has come to displace the proper component of research that should be purely exploratory.

*A literature, warning against inferential statistics, hypothesis-testing, power estimates, and the premature structuring of methodologic details and conclusions*

As introduced earlier, there is a literature of scientific philosophy that warns against unproductive or illogical practices. These warnings have been pushed aside by an effective campaign of conquest governing the formation of conventional research practice. They may be unfamiliar to many in the current audience. It might be judicious to provide an extensive, rather than brief, sampling.

Once learned, hypothesis testing based on p values led to a burgeoning biomedical literature where most reports may now be false. Discoveries providing unambiguous signs of progress are still only occasional. John Ioannidis explained how the false new "facts" are taught using complex numbers and obscure study designs, and it is difficult to know just how the results were obtained. "Research is not most appropriately represented and summarized by p-values, but, unfortunately, there is a widespread notion that medical research articles should be interpreted based only on p-values." "It can be proven that most claimed research findings are false" (14).

BF Skinner blamed Fisher and his followers for having "taught statistics in lieu of scientific method" (6).

In a popular recent book, Stuart Firestein explained that the idea that the scientific method being "one of observation, hypothesis, manipulation, further observation, and new hypothesis, performed in an endless loop…is not entirely true, because it gives the sense that this is an orderly process, which it almost never is. 'Let's

---

**"Associations" differ**

In descriptive statistics, "to have an association with" means "to be observed to exist with". The associations of a given group of subjects are independent of the features of any other group; no comparison to another group is required to establish an association. For example, fishermen as a group can be associated with the condition "holding fishing poles", even if no other group is studied.

In DS, frequencies observed in a group of subjects (e.g., the proportion of men wearing red shirts) can be compared to independent frequencies obtained from other groups (e.g., women) or from artificial results produced by random assortment (shirts of mixed colors are dropped from an imaginary helicopter). These other groups need not be highly matched to the initial group (e.g., teachers can be compared to doctors). Such comparisons are not done so as to provide the significance of the frequencies observed, but instead are for reference and illustration. DS often uses binary or categorical variables, or continuous variables having natural groupings arising from discontinuous or polymodal distributions.

Associations of interest are generally simple in DS, due to producing practical inferences that are largely intuitive. Pepe noted, however, that statistics such as the ROC and odds ratios are not intuitive to most scientists and can produce confusion unless compared to reference examples (3).

In inferential statistics, "to have an association with" means "to be associated preferentially with". The association must be supported by a low p value or other inferential statistic. To observe association in IS requires a comparison between groups.

In IS, frequencies in a group of subjects (e.g., red shirts in men, again) can be compared to another group (women) known to differ by having a nonoverlapping condition (gender). The other group can be real, or can be the frequencies expected under random assortment (the helicopter). The purpose of the comparison is to determine the significance of any differences. The matched condition or random scenario is not an independent reference, but is integral to the numerical analysis of the first group.

IS often uses continuous variables measuring small increments of value, which may be transformed to binary or ordinal variables (i.e., the variables get "dummied up") in order to search for associations. Associations (meaning preferences or differences) of interest are often subtle or complex, in which the practicality of the inferences can be obscure.

get the data, and then we can figure out the hypothesis' I have said to many a student worrying too much about how to plan an experiment…Observations, measurements, findings, and results accumulate and at some point may gel into a fact" (15) . It is the ignorance of a subject, along with an interesting means to dispel the ignorance, that drives great science.

"Science advances one funeral at a time", Max Planck is quoted as saying. Thomas Kuhn expanded on the same concept, of why a new paradigm is not immediately advanced when a hypothesis is rejected. Popular, standardizing hypotheses (paradigms) gain not merely a power far in excess of their objective worth, but an unjust near-immunity against disproof (4). In discussing new facts that ran counter to an existing paradigm, Kuhn wrote, "Assimilating a new sort of fact demands a more than additive adjustment of theory, and until that adjustment is completed – until the scientist has learned to see nature in a different way – the new fact is not quite a scientific fact at all."

David Bakan wrote, "The test of significance does not provide the information concerning psychological phenomena characteristically attributed to it; and a great deal of mischief has been associated with its use…publication practices foster the reporting of small effects in populations …[this flaw] is, in a certain sense, 'what everybody knows.'" The publication of "significant results does damage to the scientific enterprise" (13).

Distinguishing early exploratory data analysis with the follow-up use of hypothesis-testing confirmatory studies, John Tukey explained, "Unless exploratory data analysis uncovers indications, usually quantitative ones, there is likely to be nothing for confirmatory data analysis to consider…Exploratory data analysis can never be the whole story, but nothing else can serve as the foundation stone – as the first step." "Exploratory data analysis [looks] at data to see what it seems to say. It concentrates on simple arithmetic and easy-to-draw pictures…Its concern is with appearance, not with confirmation" (16).

Regarding a 1970 collation of articles condemning null-hypothesis statistical testing (The Significance Test Controversy), Jacob Cohen cites (1) contributing author Paul Meehl as having randily described the method as "a potent but sterile intellectual rake who leaves in his merry path a long train of ravished maidens but no viable scientific offspring". Schmidt and Hunter were cited by Glaser (17) as having "claimed that the use of significance testing actually retards the ongoing development of the research enterprise." Cohen agreed, writing, "I argue herein that null-hypothesis statistical testing has not only failed to support the advance of psychology as a science but also has seriously impeded it." Roger Kirk was quoted to say, "null hypothesis testing can actually impede scientific progress" (18). Charles Lambdin summarized (18), "Since the 1930s, many of our top methodologists have argued that significance tests are not conducive to science…If these arguments are sound, then the continuing popularity of significance tests in our peer-reviewed journals is at best embarrassing and at worst intellectually dishonest."

Cohen (1) provided a set of short logical puzzles illustrating illogical conclusions that can be easy to recognize as twisted when placed in familiar situations. For example, in an example paraphrased from Pollard and Richardson (19): *If a person is an American, then he is probably not a member of Congress. Yet, we know that a particular person is a member of Congress. Therefore, he is probably not an American.* Note that this ridiculous conclusion is nonetheless formally exactly the same as the following. *If $H_o$ is true, then this result (statistical significance) would probably not occur. Yet, the result has occurred. Therefore, $H_o$ is probably not true and must be formally discarded.* The dangers of null-hypothesis testing are thus made intuitive by such puzzles.

Irene Pepperberg (20) wrote, "I've begun to rethink the way we teach students to engage in scientific research. I was trained, as a chemist, to use the classic scientific method: devise a testable hypothesis, and then design an experiment to see if the hypothesis is correct or not…I've changed my mind that this is the best way to do science…First, and probably most importantly, I've learned that one often needs simply to sit and observe and learn about one's subject…Second, I've learned that truly interesting questions really often can't be reduced to a simple testable hypothesis, at least not without being somewhat absurd…Third, I've learned that the scientific community's emphasis on hypothesis-based research leads too many scientists to devise experiments to prove, rather than test, their hypotheses. Many journal submissions lack any discussion of alternative competing hypotheses."

Begley and Ellis reported that nearly 90% of pre-clinical drug studies could not be replicated (21). The early benefit of such drug candidates is typically shown by quantitative differences (not qualitative findings) obtained during hypothesis-based model-testing. I recently examined the similarly depressing realization that fewer than 1% of new cancer biomarkers enter practical use. A common and thus highly expensive cause of failure was that the "significant" result from inferential statistics was misleading. In contrast, the biomarkers that succeeded were often the markers discovered by molecular and genetic pathology and employed by clinical and surgical pathology laboratories, without depending on p values (22).

**The special problem of the p value**

The p value is at center of most applications of inferential statistics. Its major problem may be that it is not intuitive. Few investigators seem to know what it means when it is low; even fewer know what it means when it is high. Lambdin said that "the mindless ritual significance test is applied by researchers with little appreciation of its history and virtually no understanding of its actual meaning" (18).

Well published problems exist and are discussed elsewhere in this document. As summarized by Glaser (17), "the controversy involves the sole use (and misinterpretation) of the P value without taking into account other descriptive statistics, such as effect sizes and confidence intervals, statistics that provide a broader glimpse into the data analysis."

Sometimes overlooked, however, are the manipulative effects of the p value on scientific goals. The p-value mentality reinforces the desire to determine precise values. Whether 47% differs from 49% is a question demanding a p value. Tukey has been quoted as saying, "Far better an approximate answer to the right question, which is often vague, than an exact answer to the wrong question, which can always be made precise." He was echoing Aristotle's "It is the mark of an educated man to look for precision in each class of things just so far as the nature of the subject admits." An example illustrates Tukey's point. It is far more useful now to rapidly observe that a diagnostic marker fails in nearly half of the patients, than to use extensive study to determine at some later date that the precise failure rate is 47% or that it definitely fails less often than another marker.

**The special problem of conclusions**

The idea that a researcher should draw a conclusion is a concept from inferential statistics; it is not from descriptive statistics. In descriptive approaches, the data, once organized, <u>are</u> the conclusion.

According to William Rozeboom, the results of inferential statistics do not justify a decision point (i.e., a conclusion)(23). Rozeboom noted that Bayes theorem (which is an inferential technique) inherently abandons the goal of making conclusions. "The primary aim of a scientific experiment is not to precipitate decisions, but to make an appropriate adjustment in the degree to which one accepts, or believes, the hypothesis or hypotheses being tested." A confidence interval is a more suitable report of the relative confidence in a particular hypothesis. The confidence interval does not involve an arbitrary decision (i.e., a conclusion). "Insistence that published data must have the biases of the null-hypothesis decision built into the report, thus seducing the unwary reader into a perhaps highly inappropriate interpretation of the data, is a professional disservice of the first magnitude." "Its most basic error lies in mistaking the aim of a scientific investigation to be a decision, rather than a cognitive evaluation of propositions."

Tukey has also been quoted as saying, "The feeling of "Give me … the data, and I will tell you what the real answer is!" is one we must all fight against again and again, and yet again."

If investigators indeed generated true conclusions at the immediate conclusion of their studies, then research articles could be much shorter. No confirmatory studies would be justifiable. Science would not be self-correcting, it would be infallible.

Who, then, makes conclusions? Readers and clinical practitioners make conclusions. To do so, they sometimes remain patient – and use the test of time.

What can the authors and investigators properly do? They muse hypothetically, pursue proposals investigatively, suggest interpretations of data, report studies, and discover interesting things. They serve as advocates for points of view. Many, being professors, profess.

Conclusions also contain the risk of bias. When investigators make conclusions after an attempt to prove a hypothesis (such as the NIH asserted hypothesis), they have acted with bias. Such conclusions need not be trusted readily. Kuhn contrasted the suppressive action of pre-existing hypotheses, when examining data in order to reach a conclusion, with the following alternative. "The man who is ignorant of these fields, but who knows what it is to be scientific, may legitimately reach any one of a number of incompatible conclusions" (2). Retaining an open-minded legitimacy would foster innovation.

**The special problem of power estimates**

Power estimates are required in some settings. "If you plan to use inferential statistics…to analyze your evaluation results, you should first conduct a power analysis to determine what size sample you will need" (24).

In practical use, pre-test power estimation requires knowing the following.

- Your subject, well enough to have settled on a firm hypothesis and, ideally, at least one alternate hypothesis.
- The relevant sample size. Not all samples will be relevant to every statistical comparison.
- The proportion of the sample belonging to each category. These categories will then be compared by inferential methods.
- The effect size (the degree by which the dependent variable differs) that is anticipated between the categories (which are distinguished according to the independent variable).
- The practical value of various effect sizes.

- The difficulty of gathering samples, of ensuring adequate representation in each category, and of the feasibility to perform the intervention (a treatment, or an assay) on all of the samples.
- The desired alpha and beta values. These are threshold false-positive rate under the null hypothesis, and threshold false-negative results under an incorrect null hypothesis, respectively, for the study.

When a power estimate is needed, should it be performed? Maybe not, or maybe it can't - yet. Only when and if the required pre-test knowledge is present (see above) can the final power estimation be provided.

The presumed effect sizes, upon which power estimates depend, are themselves often biased, even when the effect size is based on published data from an influential scientific report (25).

For many types of research, such as molecular pathology, biochemistry, or developmental biology, an expectation or demand that all investigators must provide power estimates, prior to initiating a study, is assured to hamper selectively these particular fields. This should be self-evident; the nature of these fields is to explore areas as-yet unknown, in which the requirements for a power estimate can never be met. More subtle, however, is the destructive effects on other fields, even those relying appropriately upon inferential statistics, because many of the initial explorations in such fields will themselves depend upon employing descriptive approaches to generate their fresh ideas and momentum.

Merely to discuss power estimates can be a sign of unfamiliarity with simple statistics. Here, we can explore the underlying behaviors of group numbers, from which one can judge the utility of power estimates. Once the study population surpasses 100 samples, and certainly by 1000 samples, the relevance of the power estimate is essentially nil. At high sample numbers, the chance of discovering a meaningless or unconfirmable difference, using inferential statistics, approaches 100% for all comparisons pursued. Notably, the first published criticism of Fisher's method of determining statistical significance was in 1938. Joseph Berkson noted that p values systematically became ridiculously small when the number of samples was large; thus, decisions on "significance" were produced, even when nothing of interest was being observed (26).

As an example, let us imagine 100 samples, 40% of which belong to the first of two mutually exclusive categories. We further stipulate that the association is an exclusive one due to a causal genotype-phenotype association (e.g., if each of 40 patients having a mutation in a given gene were found to have schizophrenia, and the remaining 60 patients having a wildtype gene were not), then the two-tailed Fisher exact test would yield $p < 10^{-28}$. Even if the association were not exclusive but was barely at the threshold of clinical interest (as might be the case if 80% of the first group and 20% of the second had a particular feature), the p value would still be vanishingly low, at $<10^{-8}$. The p values would not greatly change their meaning if the proportions of the two groups began to deviate yet more from each other; this is because vanishingly low p values all convey the same meaning. Even were the first group to have only 10% of the subjects, the p value of the 80:20 difference just described would still be <.001 for the 100 samples.

The only purpose to discussing power estimates in planning or evaluating such a study exists when there is enthusiasm in finding minor differences between the groups and when there is also the opportunity to increase the sample sizes. Minor differences, such as two groups respectively having 45% and 65% prevalence of the detected feature, have little utility in molecular pathology, in developmental biology, or in many other biomedical fields. And for many initial scientific explorations, increasing the sample size may be prohibitively difficult or may not satisfy the cost/benefit consideration. Relatively unbiased, consecutively obtained, sample sets are discrete entities having a given size. They cannot be expanded on demand. It would often be preferable to explore the existing large sample set, to report the results, and to set aside the inferential studies until followup efforts to be conducted at a later date.

By this line of analysis, one may realize that in common situations existing in many fields of science, the mere discussion of power estimates can be a sign of naiveté, like a beginner without chops sitting in on a hot jam session.

When can the issue of a power estimate be raised? It should be considered when a particular hypothesis is proposed and the hypothesis envisions detecting differences (effect sizes) that are subtle, given the number of samples.

**The special problem of study planning**

Descriptive and inferential efforts differ. To plan the course of future years of inferential science requires many of the same foundations as does the procedure of power estimation. Given the requisite pieces, investigational plans tightly tied to specific hypotheses can be assembled into a formal list of procedures and rules for making decisions (conclusions). These lists readily construct a rigid, multi-year research design for future work.

In contrast, descriptive explorations do not require formal hypotheses, estimates of effect sizes, or any particular number of samples. In general, having a large number of available techniques and samples permits the greatest freedom for the anticipated exploratory efforts. Yet, even more important are a capable investigator team and a fresh subject area. An appropriate multi-year research design in exploratory efforts would list the divisions of the subject area intended for study, along with the general means for producing any unusual novel capabilities or datasets needed. Given the freshness of the field, even the methods may change monthly as these novel investigations, capabilities, and datasets find the pathways of least resistance.

As an example, the discovery of the double-helical structure of DNA by Watson and Crick had a well documented history. After long efforts and little progress, their methods were abruptly changed to seek a path of lesser resistance. This change was occasioned by a new type of data (x-ray diffraction) and advances in model-building. The change enabled rather rapid resolution of the problem.

The biomedical literature is replete with examples of notable discoveries that arose not from long-planned experiments, but from serendipity during a productive series of limited-duration exploratory efforts. This is sometimes referred to as beginner's luck, but actually reflects a predictable property in which the low-hanging fruit in a field is picked by the early, fastest-moving entrants. Such discoveries are antithetical to the delayed, routine testing of well-defined hypotheses.

It is conventional to demand detailed planning prior to funding a research effort. This creates a strong bias favoring inferential efforts. In the published guidelines governing multi-year research funding, requirements for a long-range detailed plan are common. Such a plan is suitable for well-known subjects and confirmatory efforts. Yet, these expectations come with a high price: they will systematically discriminate against exploratory examinations of fresh subject areas.

By extension of this argument, the more methodologically detailed are the expectations for long-range planning, the more biased the research-funding enterprise becomes against fast-moving explorations. Efficient exploratory research requires a subject area, a thematic goal, momentum, and investigator skills. It also requires the freedom to test and discard methods rapidly, change direction according to the lines of least resistance, and make on-the-fly decisions as to the importance of fresh data. Exploratory research is an animal body in motion. It requires a matched cognitive tool, provided by descriptive statistics.

In order to meet requirements for detailed methodological study designs, exploratory efforts must either abandon rapid innovation and descriptive statistics, or must cloak all proposals in the foreign language of inferential, hypothesis-based science.

## *Alternatives that miss the target*

Is it not that the real problem is "significance", or lack thereof? Significance is the opposite of insignificance, and innovation raises the significance of research. Let us, however, introduce some precision. In biomedicine, the significance of a discovery could refer to one of two types: 1) A discovery serving as a foundational fact on which to build a framework of understanding, such as the type of facts found in a beginning textbook of a subject, or 2) A new fact governing changes in the clinical decision-making regarding diagnosis or management. Yet, one could imagine remarkable innovation that was not immediately recognized as significant, or non-innovative confirmatory research that had high clinical significance. Worsening the priority for "significance", biomedical researchers are not philosophers. They hold to few rules regarding precise language, using the word "significance" as a leaf "uses" a stormy wind. Significance in such a linguistic wind can refer to the subject of study (disease is always a significant subject!), to the investment anticipated (to spend money and years is certainly a significant effort!), or to our nemesis, inferential statistical significance (where a low p value conveys significance by definition!). No, significance cannot be the key to innovation.

Is not the real problem "novelty", or lack thereof? Of course, novelty is fine, but to define innovation by its novelty is a tautology. And "novelty" is not as stringent a term as is "innovation". The NIH has even defined "novelty" as a grant-scoring criterion fully satisfied by the anticipation of "clinical usefulness", even when an approach is confirmatory and incremental (see below). This was an appropriate kludge, for "novelty" is irrelevant to many well established clinical research problems and has no business being a general criterion for judging biomedical research. Research has two valid directions: innovation, and its opposite. The problem is perceived unmistakably only through the lenses of statistical phyla. A study, formally designed employing power estimates, intended to reach conclusions through inferential statistics, and however useful --- is the opposite approach; innovation cannot emerge there except by serendipity.

Is not the problem "confirmatory" research? "Confirmatory data analysis" was the term used by Tukey, distinguishing "exploratory data analysis" from the followup studies initiated later. Most biomedical scientists, however, no doubt believe that their formal plans to use inferential statistics will lead to novel findings, usefulness that was unanticipated, and other valuable outcomes. And as with "significance" and "novelty", "confirmation" is an ambiguous term that can sometimes refer to a strict attempt to replicate a prior finding. We can avoid this linguistic confusion by using an operational definition instead. Innovation is opposed when a research proposal involves a compulsory study design optimized using power estimates and when inferential statistics will be used to make conclusions. If a word search, performed on a document evaluating a research direction, contains the words "hypothesis", "power", "design", or "conclusion", anxiety intensifies concerning the survival of innovation.

Finally, could it not be that the real barrier is "incrementalism"? To have "increments" requires that the research direction be previously established, a bias against innovation. And large-magnitude advances are more likely to belie true innovation than are small differences. Yet, to effectively label using the pejorative term "incrementalism" implies that scientists could agree how small is undesirably small and could recognize that even large increments might exclude innovation. If career scientists tend to agree on one thing, it is that they cannot agree on much at all. Grades of importance, of novelty, of significance, and what can be termed as incremental do not lend themselves to agreement except when re-examined comfortably later through historical hindsight. Hindsight is not an appropriate discriminator for innovative research planning, however. Instead, one could focus on the toolbox brought to the research work site. If the incrementalist's toolbox were packed with formal multiyear designs, power estimates, and plans for inferential statistics, the research will be systematically oriented towards strongly pre-established directions and likely will be incompatible with exploratory innovation. This toolbox is objectively recognizable, while judgments of incrementalism are not.

And so, the difficulty lies in more than a word or a quality. Inferential statistics for many researchers represents the bindings of their training. It is their world. They feel uncomfortable in an alternate universe; they react in alarm when they perceive that others intend to work there. During their training, typical investigators have successfully excluded the null hypothesis multiple times. Most, however, have never uncovered a really zesty innovation. Many are fully capable of doing on-the-fly exploration and would dream of having the chance. Yet, the ubiquitous research constraints of inferential statistics compel them to self-censure, to feel that it must be improper. As Kuhn explained (4), "What questions may legitimately be asked… and what techniques employed in seeking solutions? [The answers] are firmly embedded in the educational initiation that prepares and licenses the student for professional practice. Because that education is both rigorous and rigid, these answers come to exert a deep hold on the scientific mind."

To employ a different metaphor, inferential statistics is a permission slip flashed by incremental, confirmatory, hypothesis-burdened, over-designed research in search of small-magnitude differences. Given a plausible prediction of the effect size, adequate sample numbers, and a model that limits extraneous variables, its power estimates will indicate promise and convey credibility to the project. Years in advance of analyzing the observed data, the projected p values will already elicit approbation. By contrast, in a competitive environment of research funding, exploratory research, upon request, systematically fails to produce the permission slip. Publicly funded research efforts become systematically deprived of a healthy discovery pipeline… except, perhaps, when the discovery pipeline is highly focused, formally designed, administratively sanctioned, and nearly indistinguishable from hypothesis-testing research. Human genome sequencing and transcriptome profiling research was of this latter sanctioned approach.

The problem appeared to be of large magnitude, of importance, of significance. And it <u>was</u> publicly recognized, repeatedly. After decades of skirmishes, a prescient and blunt analysis was published in three documents by a notable panel of NIH-selected biomedical research experts. And what happened?

## *Policies of the Funding Source*

Based on documents of self-critique cited below, the NIH took sides in this philosophical civil war, often and explicitly choosing hypothesis testing as its sole launch vehicle for projects in both new and established scientific fields. For decades, the NIH solicitations for funding applications proffered guidelines along the lines of "This program will fund hypothesis-based initiates to study…" or, "This program aims to fund careers of young investigators who will pursue hypothesis-based inquiry…" The agency handbooks for grant applications instructed applicants in essence to "State your hypothesis, and propose your plan for establishing its validity."

**Inferential science in NIH funding decisions, a problem of logic**

Throughout scientific history, the hypothesis was a welcome tool, a friend. An expert in a subject could embark on hypothetical musings, entertaining and discarding them freely as one would collect pretty stones on a stream bank. Or, an open-ended hypothesis could guide the research productively for years.

The NIH reinvented the hypothesis. As an NIH panel reported in 2000 (27, 28), "peer review of NIH research grants has traditionally emphasized the testing of hypotheses…the practice of many NIH study sections has been to interpret it narrowly as a formal exercise in the proposal and proof of a well circumscribed idea. Under these conditions, exploratory research…and opportunities for generating new knowledge and ideas are thereby lost. Such an exclusionary insistence on hypothesis-driven research can impede the ability of NIH to accomplish its broader charge."

The NIH hypothesis thus evolved into a succinct semantic trap, a ratchet closing ever tighter. Once a subject area had been defined, the investigator was to proffer a fundamental hypothesis that was increasingly valued the more it was persuasive. Specific Aims were expected to adhere to a coherent attempt to support the hypothesis. A proposal that did not satisfactorily assure that the hypothesis could become "established" within the scope of the research plan was treated with deprecation.

In this manner, the concept of a guiding hypothesis was suppressed. An unprovable, speculative, or imprecise hypothesis became unwelcome in the scientific toolkit.

For example, imagine the following imprecise hypothetical foundation for a research project: "Tumors of the organ of Zunkerkandl had been seldom studied due to their rarity. We now have a superb collection, the first in the world. We propose that if we examine them in a variety of ways, using experts capable in tumor explorations, and taking advantage of our recently published techniques and the technical advances to emerge in the next few years, we will probably find some fresh new ideas about cancer." In the viewpoint of the NIH panel, the classical NIH system might be expected to object, "We don't see how you are going to prioritize the research. How do you plan to prove this particular hypothesis?"

The NIH "asserted hypothesis" is a distinctive one. It does not serve the same role as a classical null hypothesis, or the role of an imprecise guiding hypothesis.

The field of statistics has previously discussed the problems of the solitary hypothesis. For example, Gigerenzer warned against the "null ritual" in which only a null hypothesis is established, but not any alternative hypothesis (6). This warning, however, was directed at a true null hypothesis intended to be discarded, one that the investigators did not hold as their own hypothesis. The earlier 1938 warning from Berkson was different (26). Berkson noted that one must 1) have an alternative hypothesis and 2) be free to accept it. He wrote, "There is never any valid reason for rejection of the null hypothesis except on the willingness to embrace an alternative one".

Except in the formal setting of clinical trial planning, NIH instructions on how to propose a hypothesis typically omit the null hypothesis, the complementary alternate hypothesis (-ses), and the classical attempt to disprove the former. Instead, one is expected to proffer a positive hypothesis to which one is already firmly committed, in order to prove it true with yet more accumulated support. It is asserted in essence to be "pre-true", requiring only additional study to be proved.

In the field of rhetoric, the NIH asserted hypothesis is classified as a logical flaw termed "privileging the hypothesis". By assembling evidence towards a certain favored hypothesis, small degrees of favorable evidence become overemphasized, the evidence in favor of competing hypotheses escapes attention, and evidence potentially opposing the asserted hypothesis is not sought. That is, there is a bias towards cherry-picking both the investigations and the data.

In an epistemologic theory popularized by Karl Popper, the only manner in which to prove a hypothesis true is by deductive logic, in which all competing hypotheses are considered and disproved by empirical falsification. The interpretations obtained under either descriptive or inferential statistics are scientific theories, i.e., they meet Popper's requirement for falsifiability (29). These and other asserted hypotheses are actually well represented in the philosophical writings of Kuhn and Popper.

The NIH asserted hypothesis is typically intended to be validated by inductive logic. Inductive approaches were criticized by Popper as being non-scientific (2), "I should still contend that a principle of induction is superfluous, and that is must lead to logical inconsistencies." Inductive logic supports a hypothesis by assembling a set of facts that are consistent with the hypothesis being true, but are not necessarily inconsistent with its being untrue.

The NIH-style asserted hypothesis is generally not represented in the theoretical statistical literature, perhaps owing to its peculiar features, which include: 1) It does not rely upon a deductive falsification approach, whereby a given hypothesis is tested in an attempt to demonstrate it to be false. Instead, it is an inductive approach. 2) For there to exist a robust procedure, by which to fairly test a declared bias, is inherently illogical and self-contradictory. 3) Any flexibility of the investigator to accept an alternative hypothesis is often considered a weakness according to the NIH model for asserted hypotheses.

The cost to this logical flaw is that investments come to be made that neither will prove nor disprove the asserted hypothesis. When investigating a popular hypothesis, the research will, in the NIH model, bestow on it a patina of proof, logically unsound but now socially fixed.

**NIH intentions to reform on the issue of the hypothesis**

The NIH noted that the needs of biomedical science were not being met. The NIH Panel of Scientific Boundaries for Review was organized in 1998. In the Executive Summary of its phase 1 report of January 2000, the report addressed the cultural war on descriptive science and the pedantic methodologic detail often demanded by the reviewers (27, 28). In its paragraph on new cultural norms recommended for adoption by NIH, it included the following. Advocacy for a particular "style of research" was not to be the role of a peer reviewer. The research proposed should be judged "without undue emphasis on minor technical details". It pointedly warned, "if NIH is to accomplish its full mission, applications that propose exploratory research and methods development, in addition to those that propose hypothesis-driven research, must be judged on their potential…impact". It cautioned that an effort must be made not to discriminate "against bold new ideas." In the later detailed sections, the report again emphasized that "all research styles must be judged on their potential" and that "both hypothesis-driven and exploratory research should be judged on their merits and promise." The problem had been diagnosed, publicized, and a treatment plan recommended.

For NIH grant applications submitted after January 25, 2010, the center in charge of most initial scientific peer review (CSR) revamped the written review and scoring system to address the new priorities as part of a set of enhancements to the NIH peer review system. To downplay a prior over-emphasis on formal hypothesis-testing, the CSR in their "Advice to Investigators Submitting Clinical Research Applications" currently states an even-handed choice, "Clearly articulate the hypotheses *or* [emphasis added] objectives and specific aims of the study" (30).

**The NIH advises reform on other issues pertinent to exploratory science**

The NIH panel report also highlighted the need to downplay the recitation of details in the research plan, in which the routines of inferential statistics and conclusion-drawing are described in multiple pages. This recitation of details had served to illuminate inferential science in the best possible light, while placing exploratory approaches in a most unflattering glare during grant review.

The notices published by the NIH have also echoed the concern. Notice NOT-OD-07-014 in 2006 (31), for example, suggested, "NIH peer review could be improved by focusing less on experimental details and more on key ideas and the scientific significance of proposed projects." It predicted that an improved peer review system could benefit by "shortening the Research Plan section and focusing it more on ideas and significance".

Examples will be familiar to investigators, but publically available examples are infrequent due to privacy concerns. An illustration from 2002 was available, however. The examples of reviewer comments at CSR's website (32) include multiple instances of the reviewers suggesting specific research details, even specific reagents, often followed by coercive phrases such as "[my own personal favorite method] might be a useful addition to these studies/protocols" or "[my own personal favorite method] is not included as an experimental technique in this proposal, but could be used." The same examples contained the familiar demand for power estimates. One review described the applicant's complex set of exploratory experiments, laboriously described in detail, yet "with no mention of power calculations to assure statistical validity of these choices". Note the demand for power estimates in an exploratory study.

In the re-vampment of peer review (NIH 2008 notice NOT-OD-09-025)(33), the prior single section of the grant application termed "Background and Significance" section was re-oriented into separate sections of Significance and Novelty. A new Approach section was created to encompass the prior sections of Preliminary Data and of Research Design and Methods. The Approach section was segregated away from other sections to aid the separate scoring of the Approach criterion. Reviewers were to "give a separate score for each" criterion, which in a 2007 NIH Peer Review report (34) were designed to "rate multiple, explicit criteria individually", and were announced in the 2008 notice to have non-overlapping scorable components.

The year 2000 panel report also advised that the overall peer review evaluation of the application should pay "close attention to potential impact and the quality of the investigator without undue emphasis on minor technical details." It pointedly criticized peer reviewers, emphasizing that they "should not be in the business of designing their next experiments" (27, 28). To downplay the prior overemphasis on methodologic detail, both applicants and reviewers currently receive written instructions from CSR that the design of the Approach was to entail both the preliminary data and multiple listed categories for review, including the rationale for the design, the feasibility of the methods incorporated into the design (generally evidenced by the preliminary data, which serve as tutorials in how interpretations or conclusions are drawn from the assays), a list of studies to be performed, and the management of possible difficulties.

The Preliminary Data section was apparently incorporated into the new, shortened Approach section in order to enforce changes in what the Panel felt was "an obsession with preliminary data" among reviewers (27). The panel report instructed, "for new ideas, little or no preliminary data may be required."

Novelty was supported in the panel report where it described that "the present system tends to discourage risk-taking and to undervalue new ideas. We urge that reviewers endorse the importance of ideas that are original and have yet to be tried." Novelty was to be judged on the novel direction or, alternately for important yet incremental clinical advances, on the "usefulness" of a proposed clinical development. In context, the report indicated that the novelty could derive solely from the direction and goal, and not necessarily from the particular methods to be used (35). As with the other recommended changes, these changes were highly sympathetic to the needs of exploratory research.

The section of applications devoted to the Research Proposal was shortened from 25 to 12 pages for grants of the R01 (the major independent research grant) type. This also was intended to reduce distraction by reviewers and applicants by methodologic details and permit more emphasis on Significance and Novelty. In the past, NIH had recommended to applicants to devote about 13-16 of their 25 pages on the Research Design and Methods (NIH notice NOT-OD-06-014). With the new application format, the same NIH notice anticipated that the new methods section might occupy only half the pages allowed in new-style applications. The page-reduction would then result in a methodologically trim listing of anticipated studies in place of the prior detail, to reduce the opportunity for methodologic distraction. As stated in an NIH tutorial video introducing the new peer review system (36), a goal of the changes was "reducing the volume of technical details, and by increasing review focus on the importance of the projects". The 2007 NIH Peer Review report (34) stated the goal to be "to reduce application length to focus on impact and uniqueness/originality, placing less emphasis on standard methodological details."

**Failure to align the NIH with the recommendations**

Despite the laudable intentions of NIH reformers, the written evidence shows that administrators in the NIH's institutes did not systematically implement the Panel's year 2000 recommendations. NIH leaders of medical institutes did not issue clear guidelines to address the concerns identified. The evidence exists in the documents written by administrators and posted publicly. Advocacy continued for particular styles of research over other valid styles. For example, requirements for formalized hypothesis-testing continued to be instituted in new programs initiated by program administrators in multiple institutes.

The model application to be followed by NHLBI/NIH applicants for a K08 Mentored Clinical Scientist Career Development Award specifies that the first page of the Research Plan must be a "Statement of Hypothesis and Specific Aims" (37). The K12 Career Development Award for Clinical Oncology supports careers intended to "perform clinical oncology therapeutic research that develops and tests scientific hypotheses based on fundamental and clinical research findings; (2) design and test hypothesis-based clinical therapeutic protocols and adjunct biological analyses" (38).

The NIAID instructs applicants that "Most NIH-funded research is hypothesis driven. State your hypothesis in your abstract and Research Plan" (39). One might wonder, and what if the year 2000 panel recommendations were followed instead?

Under the re-vampment, the usage of power estimates was not revised to become more appropriate. Their use is still advised indiscriminately by the NIH system. In an NIH grant solicitation aimed to find interactions among social, behavioral, and genetic factors (40), the administrators must have resigned themselves to the sad expectation that the grantees would be energetically searching for small differences, the type of studies that led the field of psychology to jump so enthusiastically on the p value bandwagon. This grant solicitation required "power analyses to prove the sample sizes are sufficient for analysis of the interactions being studied." In other endeavors it can, however, be ambiguous semantically whether power estimates are truly required in an NIH proposal. For example, the NIH October 2011 Guidelines for NCI Program Project (PO1) Grants (41) (a large grant program) instructs PO1 applicants how to meet the requirements for the Human Subjects section. Among the five sentences of instruction is the reminder, "Power calculations justifying the number of subjects required for the proposed studies, and plans for recruitment and retention of subjects are appropriate for inclusion". In the Vertebrate Animals instructional section, a cursory three sentences includes the reminder that "power calculations justifying the number of animals required are appropriate for inclusion". Again, one wonders; might they not also be appropriate for exclusion?

The resistance to rehabilitation also extended to the peer-review process and the NIH peer-review center, CSR. Evidence appears lacking that grant review or programmatic funding decisions now place less emphasis on hypothesis testing and methodologic detail and more on significance, novelty, and momentum than before the redesign. As in the past, reviewers today remain free to weight-as-they-choose the new categories of Significance, Innovation, and Investigator(s), and are permitted to re-emphasize the rote formalities of methodologic detail.

Formal evidence for this resistance is found in the criteria for which an applicant may appeal their grant reviews. Allowable appeals must allege the review panel to have a lack of expertise or bias/conflicts of interest, or allege major errors of scientific fact (Appeals of NIH Initial Peer Review, NOT-OD-11-064)(42). The discrepancies between the long list of revamped review guidelines (2012 CSR Reviewer Guidelines and Details of Application Changes for Research Grants and Cooperative Agreements of May 20, 2010) (35, 43) show that deviations from the goals of the Panel's report (deviations such as an imposed requirement for formally testing an NIH-style asserted hypothesis, or requiring power estimates for early exploratory studies) were not adopted in the revampment as grounds for appeal. Reviewers are also still permitted to view the detailed research plan as a determinant of the proposal's significance and novelty, rather than as a separate criterion for evaluation.

As documented above in examples, grant review criteria previously could exclude exploratory proposals by requiring hypotheses, power estimates, and prescriptive study design. After the revampment, criteria biased in favor of the incremental discovery of pre-conceived distinctions continued to be acceptable in grant review. Casual observers may not be aware of the blatant openness of the philosophical battles over power estimates and inferential statistics waged during grant review at the NIH, and thus may not appreciate its effects to limit exploratory science in under-studied areas of biomedical sciences. For example, in a biomedical grant proposal reviewed in 2012 by NIH's CSR (44), the applicant proposed studies of a subject (the possible function of a novel gene), in which the statistical sections of the application expressly proposed to employ descriptive statistics to the exclusion of inferential statistics, explaining, "We should note that we are not interested in having a high power to detect subtle differences… we are looking for fundamental distinctions and natural separations of the data." The official review of the proposal, in a document issued by CSR, protested this approach, requesting new "power calculations performed" and stipulated a new research goal "to obtain statistically significant results such that even minimal associations will be made."

The NINDS warns applicants of the "Eight basic questions reviewers ask", one of which is "Is the hypothesis valid and have you presented evidence supporting it?" (45). Yet, decades of writers have suggested, might not the testing of a valid hypothesis itself be criticized? Why test valid hypotheses, when more novel areas beckon?

Over the past decade, even the initiatives intended to fund descriptive and anatomic science (such as the -omic anatomy of diseases) have been couched in the language of inferential statistics. Certain administratively sanctioned categories of descriptive science became renamed "hypothesis-generating science" to stand shoulder-to-shoulder with the "hypothesis-testing" expected of other approaches. In the eyes of the new NIH review system, an intriguing exploratory excursion must still be clad in the robes of formal and methodologically-encased hypothesis-spun cloth.

In such ways, the statistical battles of the 1930s revisit the current day, but with an unfair institutional advantage favoring one side.

In summary, despite the revampment at CSR, the NIH today often overtly leans towards hypothesis-testing science and inferential statistics. This stance can displace exploratory science and the attendant descriptive statistical approaches. This tendency is most harmful to under-served research subjects in which the exploratory data do not yet exist to support confirmatory study designs. From the panel's statements, and from the perspective of a broad scientific literature, the continued deference towards inferential statistics, power estimates, and detailed study design continues to shortchange biomedical progress.

## *Possible benefits from reform*

Through displacement, systematic preferences for inferential statistics can cull and replace valuable components of a research program, whether local, national, or international. A forceful reversal is advised for the perniciously harmful encroachment of inferential formalism into our research goals, into peer-review, into funding decisions, and into study designs. Low-impact, low-efficiency, low-speed research programs could be re-oriented into more fruitful efforts.

What level of improvement could be expected in research productivity from such re-orienting? Measures of improvement would depend on the productivity cost currently being borne due to excessive inferential formalism and on the ability to redirect efforts into exploratory research based upon descriptive statistics. One would need to consider assigning:

- An estimate of the current fraction of efforts devoted to exploratory approaches
- An estimate of the current fraction of high-impact clinically valuable research derived specifically from exploratory approaches
- The possible degree by which efforts could be re-purposed towards research grounded in exploratory efforts

Such numbers are not known, but estimates will explore these concepts. Let us stipulate, for sake of argument, that current efforts towards descriptive approaches and qualitative discoveries constitute a minority of the research portfolio and yet yield the majority of high-impact discoveries. Under these premises, the rate of high-impact discoveries would rise considerably if the formalities of hypothesis-based inferential science were loosened. This rise in output would manifest over some time, as investigators began first to engage in, then harvest from the new research directions and practices.

## *Recommendations*

We should re-orient biomedical research towards a balance of approaches in which descriptive statistics and qualitative differences play prominent roles.

The above history reveals that the constraints on innovative research are not merely matters of judgment and preference. The cited documents illustrate that a recognizable subset of administrative procedures, announcements, grant scoring criteria, and appeal criteria operate to set the conditions. Operational definitions can measure and document these policies. The resulting metrics would permit agencies to monitor the biases towards inferential statistics, power estimates, and prescriptive study design in most externally funded biomedical research. Using these same analytical tools, we should argue, we produce tangible means to overcome the institutional biases against exploratory research.

But do inferential statistics constrain research, or do people constrain research? Any corrective actions should be directed to the influential persons. Suggestions for remediation follow.

- In the teaching of biomedical statistics, <u>to devote</u> adequate weight to the role of descriptive statistics. Among three general biostatistical texts on my bookshelf, none devoted more than a page or two to the subject. Inferential statistics, no doubt, deserves more instructional time and textbook space owing to its greater complexity and to its role in therapeutic development. Insufficient attention, however, is given to simple associations, their clinical impact and their importance to practical daily decision-making.

- <u>To instruct</u> researchers on the appropriate role of each type of statistical analysis. This must go beyond merely warning authors about the danger of the p value. The boxed text suggests goals for such instruction.

- <u>To instruct</u> authors, editors, and peer reviewers that writing about a hypothesis mandates that the merits of the alternative hypotheses must also be explained proportionately.

- <u>To instruct</u> decision-makers at funding agencies in the inherent distinctions between descriptive and inferential discoveries and how an improved recognition of their distinctive differences would promote overall research impact.

- <u>To study and report</u> summary statistics as to what fraction of grant funds are awarded for each of the two statistical phyla: descriptive and inferential statistics

- In the funding preferences, <u>to encourage</u> using administrative solicitations the research proposals that respect the special purposes of the different statistical phyla.

- <u>To make</u> paramount the criteria of significance, innovation, investigator capabilities, timeliness, and the momentum in assessing research proposals, so that funding will encourage innovation.

- Leaders in peer review, such as editors and review administrators, should be made conversant with the major branches of statistics and their separate purposes, so as <u>to ensure</u> fair and goal-appropriate review, <u>to identify</u> errant reviewers for education or substitution, and <u>to cull out</u> inappropriate reviews.

- <u>To instruct</u> peer-reviewers that to judge descriptive approaches by the rules of inferential formalism, or to judge inferential approaches by the rules of the purely exploratory sciences, is banned. Enforcement efforts should be active.

- <u>To institute</u> consistent computer review of peer review documents to cull out for manual examination the possible pejorative uses of word roots such as "power" (as in underpowered, no power estimate), "hypothe" (as in weak hypothesis, not hypothesis-driven), "conclu" (as in rules for making conclusions, plan for concluding significance), "significan", etc.. Not all uses would be inappropriate; for therapeutic trials, such an examination could be cursory. In the field of molecular pathology, such examination would be of paramount importance to ensure fair review and health of the field.

- <u>To encourage</u> appeals of grant peer review for inappropriate scientific scope. Proposals to perform exploratory science in search of qualitative discoveries using the toolbox of descriptive statistics must not be reviewed according to a restricted scope, one in which inferential science, power estimates, incremental small differences, and excessive formal design are expected.

While implementing these recommendations, certain broad concepts need to be kept in mind. Significance refers to the downstream impact anticipated should the mechanics of the proposal succeed, as adjusted by the Bayes' prior probability of success. Neither the novelty nor the significance of an exploratory proposal depends on the efficiency of the methodological approach and perhaps not even on the proposed breadth of the study.

When something novel is being pursued, it would be absurd to expect an accurate multi-year methodologic plan. Indeed, the favored methods, when undertaking descriptive approaches, may change yearly or even week-by-week as technologies improve and as an investigator pursues new findings on-the-fly. The methods in inferential approaches, by contrast, are often much more rigid, given the necessity for an established plan that is adequately specific to support a power estimate and protocol-guided recruitment of patients or samples.

In short, descriptive and inferential approaches to research do differ. Their infant science projects should be raised differently. Reforms are overdue.